\documentclass[12pt]{article}
\usepackage{graphicx}
\usepackage{geometry}
\usepackage{graphicx}
\usepackage{enumerate}
\begin{document}
\begin{center}
{\large {\bf Heliosheath: Diffusion Entropy Analysis\\ and Nonextensivity q-Triplet}}\\[0.5cm]

{\bf A. Haubold$^1$, H.J. Haubold$^{2,3}$, and D. Kumar$^3$}\\
\medskip
\small{$^1$ Department of Computer Science, Columbia University, New York, NY, USA (ah297@columbia.edu)\\[0.2cm]
$^2$ Office for Outer Space Affairs, United Nations, P.O. Box 500, Vienna International Centre, A-1400 Vienna, Austria (hans.haubold@unvienna.org)\\[0.2cm]
$^3$ Centre for Mathematical Sciences Pala Campus, Arunapuram P.O., Pala, Kerala-686574, India (dilipkumar.cms@gmail.com)\\[0.2cm]}
\end{center}
\bigskip
\noindent
{\bf Abstract.} In this paper we investigate the scaling behavior, based on Diffusion Entropy Analysis and Standard Deviation Analysis, of the magnetic field strength fluctuations recorded by Voyager-I in the heliosphere.  The Voyager-I data set exhibits scaling behavior and may follow L\'{e}vy-type probability distribution. A general fractional-order spatial and temporal diffusion model could be utilized for the interpretation of this L\'{e}vy-type behavior in comparison to Gaussian behavior. This result confirms earlier studies of scaling behavior of the heliospheric magnetic field strength fluctuations based on non-extensive statistical mechanics leading to the determination of the nonextensivity $q$-triplet.

\section{Introduction}

The atmosphere of the Sun beyond a few solar radii (heliosphere) is fully ionized plasma expanding at supersonic speeds, carrying solar magnetic fields with it [1]. This solar wind is a driven non-linear non-equilibrium system. The Sun injects matter, momentum, energy and magnetic fields into the heliosphere in a highly variable manner. Fluctuations of the magnetic field strength in the solar wind may have fractal and multifractal scaling structure over a large range of scales in the region from 1 to 85 AU. This type of scaling represents a hierarchical structure in phase space as represented in nonextensive statistical mechanics, in contrast to the uniformly occupied phase space of Boltzmann-Gibbs statistical mechanics [2].\\

The spacecraft Voyager-I crossed the termination shock at a distance of 40 AU (1989) and 85 AU (2002) from the Sun and moved through the heliosheath toward the interstellar medium. The distributions of magnetic field strength fluctuations observed in the heliosheath are Gaussian over a wide range of time scales. The probability density of changes in the magnetic field strength on short-time scales exhibit non-Gaussian behavior while on long-time scales the Gaussian assumption has been confirmed [3,4]. This suggests that the inner heliosheath is not in statistical equilibrium.\\

For the basic quantities (e.g., sensitivity to the initial conditions, relaxation towards equilibrium of correlation functions, equilibrium distribution of energies) of systems described by Boltzmann-Gibbs statistical mechanics, the exponential function emerges ubiquitously as the adequate one. For non-equilibrium situation (e.g., non-equilibrium of energies, other chaotic situations) a new and more general density function is needed. In order to take care of these non-equilibrium, unstable or chaotic situations the exponential type function is replaced by the q-exponential function for systems described by nonextensive statistical mechanics [5,6] or the entropic pathways [13,15,16]. The question appears about what is the value of $q$ and which precise physical meaning can be ascribed to $q$. The dynamical scenario for nonextensive statistical mechanics is an appropriate quantifier of nonextensivity in terms of a $q$-triplet. Such a $q$-triplet was theoretically predicted by Tsallis [5,6]. Accordingly, a stationary state is characterized by a triplet of $q$-values $(q_{stat}, q_{sen}, q_{rel}) \neq (1,1,1)$, where $q_{stat} > 1, q_{sen} < 1, \mbox{and}\; q_{rel }> 1.$ Subsequently, the analysis of experimental and observational data in terms of non-extensive statistical mechanics led to the discovery of q-triplets for the magnetic field strength fluctuations [3,4], temperature fluctuations of the microwave background radiation [7], solar magnetic activity [8], and stratospheric ozone layer variability [9]. The q-triplet for the magnetic field strength fluctuations of the solar wind is shown in Figure 1.\\

In this paper we analyze the Voyager-I magnetic field strength recordings with standard deviation analysis (SDA) and diffusion entropy analysis (DEA) methods focusing on the probability density function realized in the data set [11]. A similar analysis had been undertaken for the solar neutrino flux as recorded in the Super-Kamiokande-I experiment [10].
\begin{figure}[htp]
\resizebox{15cm}{!}{\includegraphics{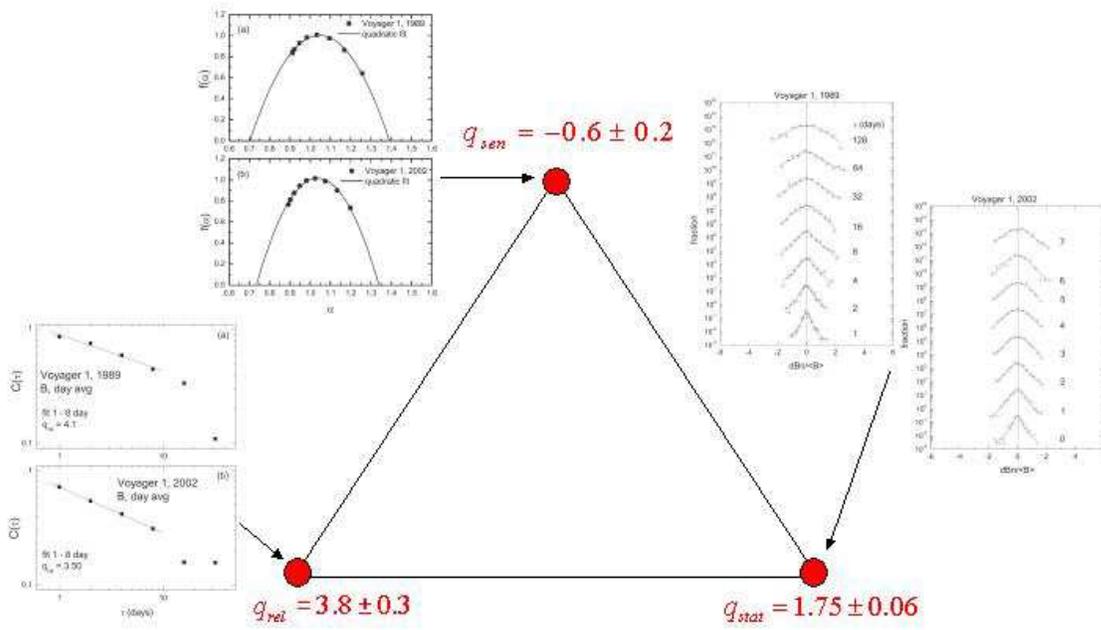}}\\
\caption{The q-triplet as measured from the magnetic field strength fluctuations of the solar wind. The three sets of curves correspond to daily averages of the data from 1989 (40 AU) and from 2002 (85 AU) by NASA`s Voyager-I spacecraft [3,4,5].}\par
\end{figure}\par
\begin{figure}[htp]
\resizebox{15cm}{!}{\includegraphics{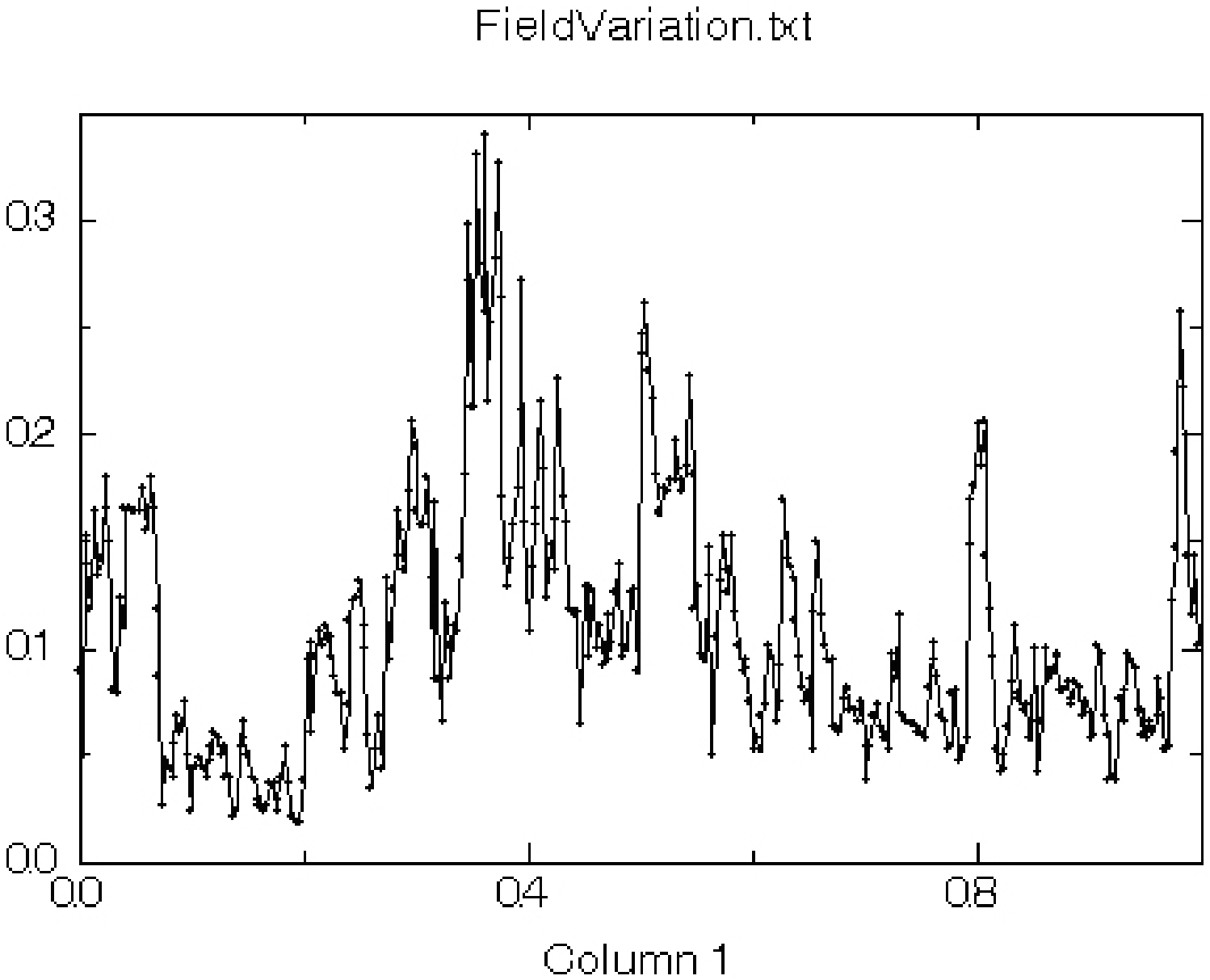}}\\
\caption{The figure shows the daily averages of the magnetic field strength fluctuations versus time $t$ measured by Voyager-I during 1989 (45 AU).}
\end{figure}

\section{Methods of Analysis: SDA and DEA}

To study the scaling behavior of the Voyager-I magnetic field strength data (Figure 2) we make use of two complementary scaling analysis methods: The Standard Deviation Analysis (SDA) and the Diffusion Entropy Analysis (DEA). The purpose for utilizing the two methods to analyze the scaling properties of the time series is to discriminate the stochastic nature of the data sets: Gaussian or L\'{e}vy-type.  Such methods have been developed in great detail by Scafetta [11, 12] emphasizing the fact that the SDA is based on the analysis of the variance of the data while the DEA directly measures the probability density function $p(x,t)$ of the data.\par
\bigskip
\noindent

The DEA perceives the numbers in a time series as the trajectory of a diffusion process. In the following $x$ denotes the diffusion variable of such a process. The scaling property of $p(x,t)$ takes the form
\begin{equation}
p(x,t)=\frac{1}{t^\delta}F(\frac{x}{t^\delta}).
\end{equation}

Scaling laws reflect underlying physical principles that are independent of detailed dynamics of particular models. Scale invariance is widespread in natural systems [17].

In the SDA one examines the scaling properties of the second moment of the diffusion process generated by a time series. The standard deviation $D(t)$ of the variable $x$ is
\begin{equation}
D(t)=(<x^2,t>-<x,t>^2)^{1/2}\;\;\propto\;t^H.
\end{equation}

The Hurst exponent $H$ is interpreted as the scaling exponent being evaluated from the gradient of the fitted straight line in the log-log plot of the entropy $S(t)$ over time $t$.
The DEA determines the scaling exponent $\delta$ evaluated through the Shannon entropy $S(t)$ of the diffusion process generated by the variation of the time series. The probability density function $p(x,t)$ is determined by means of the sub-trajectories
\begin{equation}
X_n(t)=\sum^l_{i=0}\xi_{i+n},\;\;n=0,1,\ldots
\end{equation}

If the scaling condition in (1) is valid, the corresponding entropy increases over time as
\begin{equation}
S(t)=-\int^{+\infty}_{-\infty}{\rm d}x ~p(x,t)\ln[p(x,t)].
\end{equation}
Using the scaling condition in (1), we obtain
\begin{equation}
S(t)=A+\delta \ln(t).
\end{equation}

With
\begin{equation}
A=-\int^{+\infty}_{-\infty}{\rm d}y ~F(y)~\ln[F(y)]=\mbox{const.,}
\end{equation}

where $y = \frac{x}{t^\delta}$. Eq. (5) indicates that in the case of a diffusion process with a scaling probability density, its entropy $S(t)$ increases linearly with $\ln(t)$. The scaling exponent $\delta$ is being determined from the gradient of the fitted straight line in the linear-log plot of $S(t)$ over time $t$.
For fractional Brownian motion the scaling exponent $\delta$ coincides with the Hurst exponent $H$. For random noise with finite variance, the probability density function $p(x,t)$ will converge, according to the central limit theorem, to a Gaussian distribution with $H = \delta = 0.5$. If the Hurst exponent $H$ is different from the scaling exponent $\delta$, the scaling represents anomalous behavior. The diffusion process characterized by L\'{e}vy  flights belongs to anomalous diffusion. In this specific case (1) is still valid but the variance is not finite and the variance scaling exponent can not be defined. A second type of anomalous diffusion is due to L\'{e}vy walk based on a generalization of the central limit theorem. In this case of anomalous diffusion, the second moment is finite and the exponents $H$ and $\delta$ obey the relation
\begin{equation}
\delta=\frac{1}{3-2H}.
\end{equation}

The results of applying SDA and DEA to the time series of Voyager-I magnetic field strength fluctuations are shown in Figure 3 and Figure 4. These plots are fitted to (2) and (5), respectively, yielding the scaling exponents $H = 0.90$ and $\delta = 0.83$ for the magnetic field strength fluctuations. Standard deviation scaling exponent $H$ and diffusion entropy scaling exponent $\delta$ are markedly larger than $0.5$ which is taken as an indication to follow the L\'{e}vy-type probability density function. This result supports the conclusion for nonextensivity of the data set as derived in [3,4].

\begin{figure}
\resizebox{15cm}{!}{\includegraphics{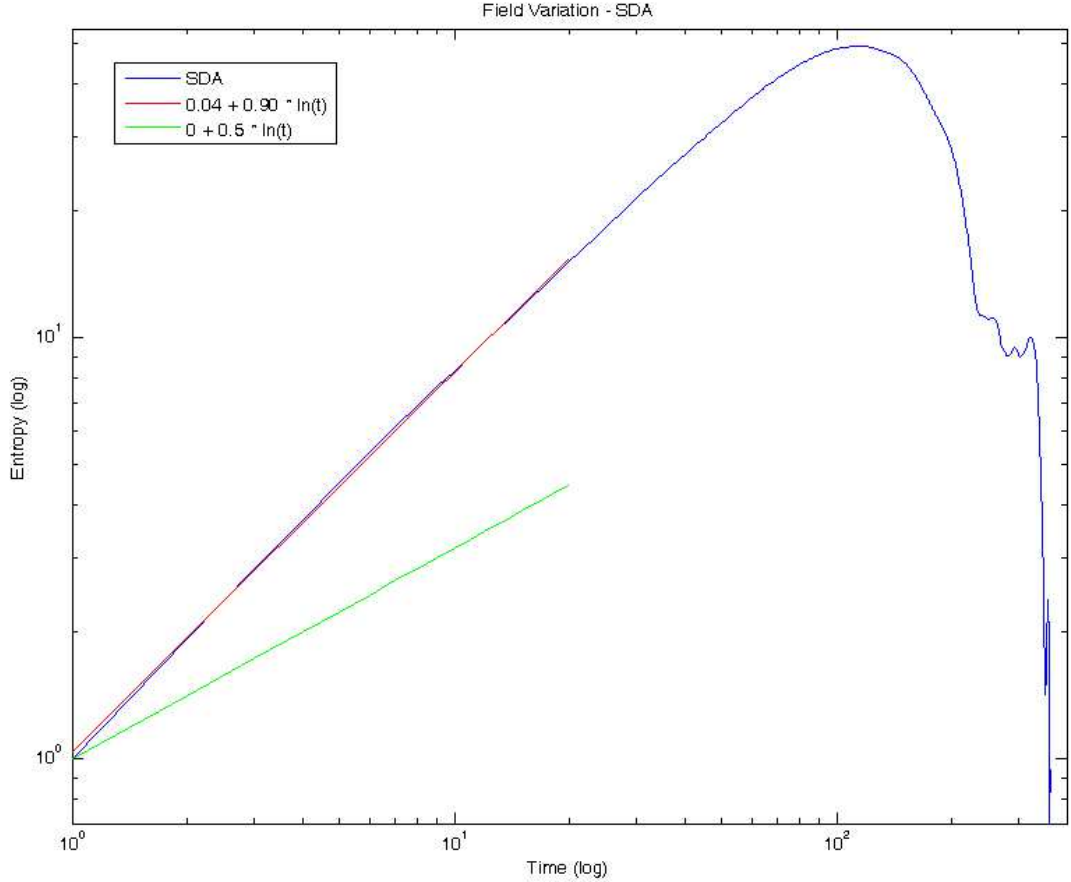}}\\
\caption{Numerical standard deviation analysis results in blue with a fitting line in green for $H = 0.5$ and in red for $H = 0.90$, respectively.}
\end{figure}

\begin{figure}
\resizebox{15cm}{!}{\includegraphics{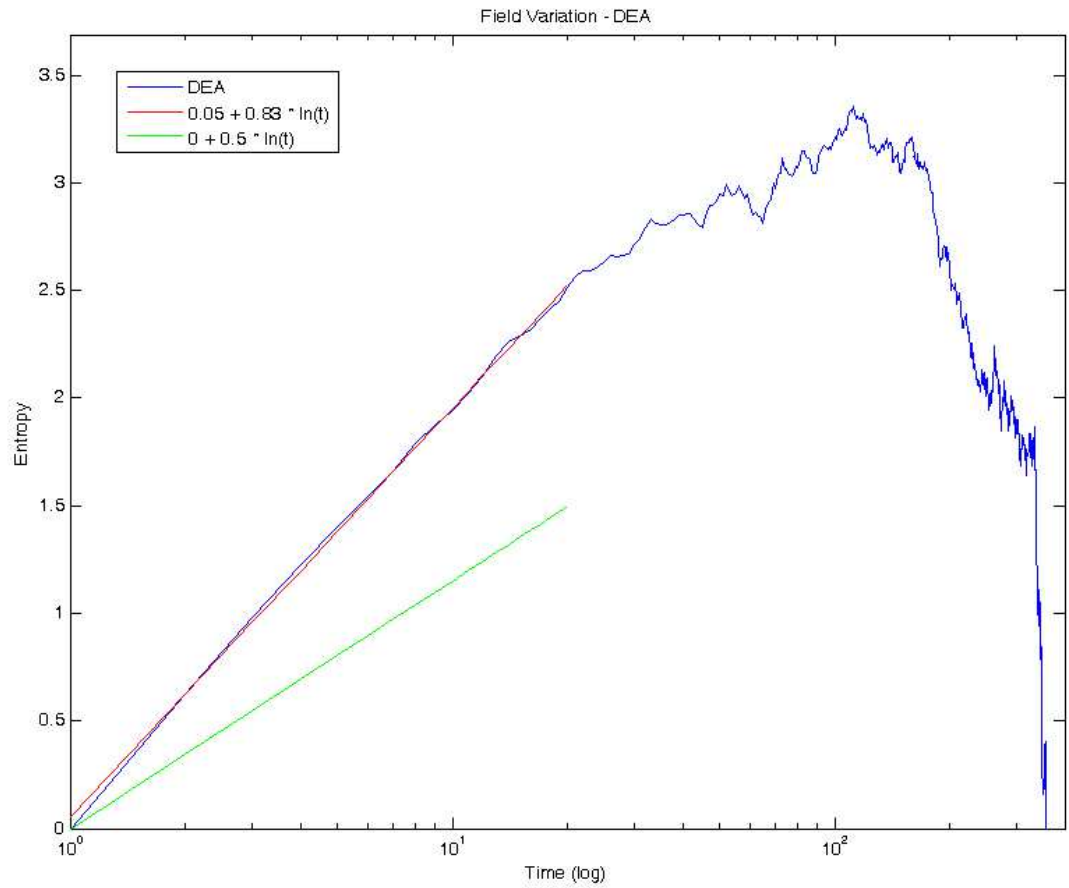}}\\
\caption{Numerical diffusion entropy analysis results in blue with a fitting line in green for $\delta = 0.5$ and in read for $H = 0.83$, respectively.}
\end{figure}

\section{Preliminary Conclusions}

Initial ideas to model diffusion processes showing the scaling of probability density functions in Figures 3 and 4 we consider the following diffusion model with fractional-order spatial and temporal derivatives [13,14,15,16] 
\begin{equation}
_0D_t^\beta N(x,t)=\eta\;_xD_\theta^\alpha N(x,t),
\end{equation}
with the initial conditions $_0D_t^{\beta-1}N(x,0)=\sigma(x), 0\leq\beta\leq 1, \lim_{x\to \pm\infty} N(x,t)=0,$ where $\eta$ is a diffusion constant; $\eta,t>0,x\in R; \alpha, \theta, \beta$  are real parameters with the constraints
$$0<\alpha \leq 2, |\theta| \leq min(\alpha, 2-\alpha),$$
and $\delta(x)$ is the Dirac-delta function. Then for the fundamental solution of (8) with initial conditions, there holds the formula
\begin{equation}
N(x,t)=\frac{t^{\beta-1}}{\alpha|x|}H^{2,1}_{3,3}\left[\frac{|x|}{(\eta t^\beta)^{1/\alpha}}\left|^{(1,1/\alpha), (\beta,\beta/\alpha), (1, \rho)}_{(1,1/\alpha), (1,1), (1,\rho)}\right.\right], \alpha>0
\end{equation}
where $\rho=\frac{\alpha-\theta}{2\alpha}.$
\medskip
\noindent

The following special cases of (8) are of special interest for fractional diffusion models:

(i)  For $\alpha=\beta$,  the corresponding solution of (8), denoted by $N_\alpha^\theta$,  we call as the neutral fractional diffusion, which can be expressed in terms of the H-function as given below and can be defined for $x>0$:\par
\medskip
\noindent
Neutral fractional diffusion: $0<\alpha=\beta<2; \theta\leq \min \left\{\alpha,2-\alpha\right\},$
\begin{equation}
N_\alpha^\theta(x)=\frac{t^{\alpha-1}}{\alpha|x|}H^{2,1}_{3,3}
\left[\frac{|x|}{t\eta^{1/\alpha}}\left|^{
(1,1/\alpha),(\alpha,1), (1,\rho)}_{(1,1/\alpha),(1,1), (1,\rho)}\right]\right.,\; \rho=\frac{\alpha-\theta}{2\alpha}.
\end{equation}		
(ii) When $\beta=1,0<\alpha\leq2;\theta\leq \min\left\{\alpha, 2-\alpha\right\}$, then (8) reduces to the space-fractional diffusion equation, which is the fundamental solution of the following space-time fractional diffusion model:
\begin{equation}
\frac{\partial N(x,t)}{\partial t}= \eta\;_xD_\theta^\alpha N(x,t), \eta>0, x\in R,
\end{equation}
with the initial conditions  $N (x,t=0) = \sigma(x), \displaystyle {\lim_{x\to\pm\infty}} N(x,t)=0,$ where $\eta$ is a diffusion constant  and $\sigma(x)$ is the Dirac-delta function. Hence for the solution of (8) there holds the formula
\begin{equation}
L_\alpha^\theta(x)=\frac{1}{\alpha(\eta t)^{1/\alpha}}\;H^{1,1}_{2,2}\left[\frac{(\eta t)^{1/\alpha}}{|x|}\left|^{(1,1),(\rho, \rho)}_{(\frac{1}{\alpha},\frac{1}{\alpha}),(\rho, \rho)}\right]\right.,\;0<\alpha<1, |\theta|\leq \alpha,
\end{equation}
where $\rho=\frac{\alpha-\theta}{2\alpha}$. The density represented by the above expression is known as $\alpha$-stable L\'{e}vy density. Another form of this density is given by
     \begin{equation}
L_\alpha^\theta (x)=\frac{1}{\alpha(\eta t)^{1/\alpha}}\;H^{1,1}_{2,2}\left[\frac{|x|}{(\eta t)^{1/\alpha}}\left|^{(1-\frac{1}{\alpha},\frac{1}{\alpha}), (1-\rho, \rho)}_{(0,1),(1-\rho, \rho)}\right.\right], 1<\alpha< 2, |\theta|\leq 2-\alpha.
\end{equation}
(iii) Next, if we take $\alpha=2,0<\beta<2;\theta =0$,  then we obtain the time-fractional diffusion,  which is governed by the following time-fractional diffusion model:
\begin{equation}
\frac{\partial ^\beta N(x,t)}{\partial t^\beta} = \eta \frac{\partial^2}{\partial x^2} N(x,t), \eta>0, x\in R,\; 0<\beta\leq 2,
\end{equation}
with the initial conditions $_0D_t^{\beta-1} N(x,0)= \sigma(x), _0D_t^{\beta-2} N(x,0)=0,\; \mbox{for}\; x \in r, \\ 
\lim_{x\to\pm\infty} N(x,t)=0$,
where $\eta$  is a diffusion constant  and $\sigma(x)$ is the Dirac-delta function, whose fundamental solution is given by the equation
\begin{equation}
N(x,t)=\frac{t^{\beta-1}}{2|x|}\;H^{1,0}_{1,1}\left[\frac{|x|}{(\eta t^\beta)^{1/2}}\left|^{(\beta, \beta/2)}_{(1,1)}\right.\right].
\end{equation}
(iv) If we set $\alpha=2, \beta=1$ and $\theta\rightarrow 0$, then for the fundamental solution of the standard diffusion equation
\begin{equation}
\frac{\partial}{\partial t}N(x,t)=\eta\frac{\partial^2}{\partial x^2}N(x,t),
\end{equation}
with initial condition
\begin{equation}
N(x,t=0) = \sigma(x), \lim_{x\to \pm\infty} N(x,t)=0,
\end{equation}
there holds  the formula
\begin{equation}
N(x,t)=\frac{1}{2|x|}H^{1,0}_{1,1}\left[\frac{|x|}{\eta^{1/2}t^{1/2}}
\left|^{(1,1/2)}_{(1,1)}
\right.\right]=(4\pi \eta t)^{-1/2} \exp[-\frac{|x|^2}{4\eta t}],
\end{equation}
which is the classical Gaussian density.\par

In conclusion we have studied the time series of Voyager-I magnetic field strength variation and discovered scaling behavior of the variation in the dataset. Using two scaling methods, we have evaluated SDA and DEA scaling exponents. The scaling exponent for the Voyager-I data indicates that the data set exhibits L\'{e}vy-type behavior. A general fractional-order spatial and temporal diffusion model could be utilized for the interpretation of this L\'{e}vy-type behavior in comparison to Gaussian behavior. \par
\medskip
\noindent
{\bf Acknowledgments}. We profoundly appreciate enlightening discussions with A.M. mathai, Director of the Centre for Mathematical Sciences. We are grateful to Leonard F. Burlaga and Constantino Tsallis for providing us the Voyager-I data and insightful advice on their interpretation in terms of nonextensive statistical mechanics. Nicola Scafetta generously made available to us the C++ code for SDA and DEA. We also thank the Department of Science and Technology, Government of India, New Delhi, for the financial assistance for this work
 under project No. SR/S4/MS:287/05, and the Centre for Mathematical Sciences
 for providing all facilities.\par
\bigskip
\noindent
{\bf References}\par
\medskip
\noindent
[1] Burlaga, L.F.: 1995, Interplanetary Magnetohydrodynamics, International Series on Astronomy and Astrophysics, Oxford University Press, New York and Oxford.\par
\smallskip
\noindent
[2] Tsallis, C.: 2009, Introduction to Nonextensive Statistical Mechanics: Approaching a Complex World, Springer, New York.\par
\smallskip
\noindent
[3] Burlaga, L.F. and Vinas, A.F.: 2005, Triangle for the entropic index q of non-extensive statistical mechanics observed by Voyager 1 in the distant heliosphere, Physica A, 356, 375-384.\par
\smallskip
\noindent
[4] Burlaga, L.F., Vinas, A.F., Ness, N.F., and Acuna, M.H.: 2006, Tsallis statistics of the magnetic field in the heliosheath, The Astrophysical Journal, 644, L83-L85.\par
\smallskip
\noindent
[5] Tsallis, C.: 2009, Nonadditive entropy and nonextensive statistical mechanics – An overview after 20 years, Brazilian Journal of Physics, 39, 337-356.\par
\smallskip
\noindent
[6] Tsallis, C.: 2004, Dynamical scenario for nonextensive statistical mechanics, Physica A, 340, 1-10.\par
\smallskip
\noindent
[7] Bernui, A., Tsallis, C., and Villela, T.: 2006, Temperature fluctuations of the cosmic microwave background radiation: A case of non-extensivity?, Physics Letters, A 356, 426-430.\par
\smallskip
\noindent
[8] De Freitas, D.B. and De Medeiros, J.R.: 2009, Nonextensivity in the solar magnetic activity during the increasing phase of solar cycle 23, European Physics Letters, 88, 19001-p1 - 19001-p6.\par
\smallskip
\noindent
[9] Ferri, G.L., Reynoso Savio, M.F., and Plastino, A.: 2010, Tsallis' $q$-triplet and the ozone layer, Physica A, 389, 1829-1833.\par
\smallskip
\noindent
[10] Haubold, A., Haubold, H.J., and Kumar, D.: 2012, Solar neutrino records: Gauss or non-Gauss is the question, arXiv: 1201.1549v1.\par
\smallskip
\noindent
[11] Scafetta, N.:  2010, Fractal and Diffusion Entropy Analysis of Time Series: Theory, concepts, applications and computer codes for studying fractal noises and Levy walk signals, VDM verlag Dr. Mueller, Saarbruecken.\par
\smallskip
\noindent
[12] Scafetta, N.: 2004, in Gell-Mann, M. and Tsallis, C. (Eds.), 2004, Nonextensive Entropy: Interdisciplinary Applications, Santa Fe Institute Studies in the Sciences of Complexity, Oxford University Press, Oxford and New York.\par
\smallskip
\noindent
[13] Mathai, A.M., Saxena, R.K., and Haubold, H.J.:  2010, The H-Function: Theory and Applications, Springer, New York.\par
\smallskip
\noindent
[14] Haubold, H.J., Mathai, A.M., and Saxena, R.K.: 2011, Further solutions of fractional reaction-diffusion equations in terms of the H-function, Journal of Computational and Applied Mathematics, 235, 1311-1316.\par
\smallskip
\noindent
[15] Mathai, A.M. and Haubold, H.J.: 2011, A pathway from Bayesian statistical analysis to superstatistics, Applied Mathematics and Computation, 218, 799-804.\par
\smallskip
\noindent
[16] Mathai, A.M. and Haubold, H.J.: 2011, Matrix-variate statistical distributions and fractional calculus, Fractional Calculus and Applied Analysis, 14, 138-155.\par
\smallskip
\noindent
[17] Mandelbrot, B.B.: 1983, The Fractal Geometry of Nature, W.H. Freeman and Company, New York.\par
\end{document}